\documentclass[12pt]{article}
 \usepackage[dvips]{graphicx}
\begin{document}
\title{On p-adic Stochastic Dynamics, Supersymmetry and the Riemann Conjecture}
\bigskip

\author{ Carlos Castro \thanks{Center for Theoretical Studies of Physical Systems,Clark Atlanta
University, Atlanta, GA. 30314; e-mail: castro@ctsps.cau.edu}}
\date{}
\maketitle
\centerline { (dedicated to the memory of Michael Conrad) }
\bigskip
\begin{abstract}
We construct ( assuming the quantum inverse scattering problem has a
solution ) the operator that yields the zeroes of the Riemman zeta function
by defining explicitly the supersymmetric quantum mechanical model
( SUSY QM ) associated with the $p$-adic stochastic dynamics of a particle
undergoing a Brownian random walk . The zig-zagging occurs after collisions
with an infinite array of scattering centers that $fluctuate ~randomly$.
Arguments are given to show that this physical
system can be modeled as the scattering of the particle about the
infinite locations of the prime numbers positions.  We are able then to
reformulate such $p$-adic stochastic process, that has an underlying hidden
Parisi-Sourlas supersymmetry, as the $effective$ motion of a particle in a
potential which can be expanded in terms of an infinite collection of 
$p$-adic harmonic oscillators
with fundamental (Wick-rotated imaginary ) frequencies $ \omega_p = i log~p$
($ p $ is a prime ) and whose harmonics are $ \omega_{p, n} = i log ~ p^n$.
The $p$-adic harmonic oscillator potential allow us to determine a
one-to-one correspondence between the amplitudes of oscillations $a_n$
( and phases ) with the imaginary parts of the zeroes of zeta $ \lambda_n$ , after
solving the inverse scattering problem.
\end{abstract}
\bigskip
The Riemann conjecture that the nontrivial zeroes of the zeta function lie
on the vertical line $z = 1/2 + i y $ of the complex plane remains one of
the most important unsolved problems in pure Mathematics. Hilbert and Polya
suggested long ago that the zeroes of zeta could have an spectral
interpretation in terms of the eigenvalues of a suitable self adjoint
trace class linear differential operator. Finding such operator, if it exists,
will be tantamount to  proving the Riemann conjecture. There is a related
analogy with the Laplace-Beltrami operator in the Hyperbolic plane ,
a surface of constant negative curvature. The motion of a billiard in such
surfaces is a typical example of classical chaotic motion. The Selberg
zeta function associated with such Laplace-Beltrami operator
admits zeroes which can be related to the energy eigenvalues of such
operator.
Since the zeroes of the Riemann zeta function are  deeply connected with
the distribution of primes it has been suggested by many authors
 that the spectral properties of the zeroes of zeta may be associated with
the statistical fluctuations of the energy levels of a quantum system whose 
underlying classical dynamics are chaotic.  Random matrix theory  [1] has numerous physical applications
in all branches of science.  Connections bewteen the asymptotic distribution of primes $x/log x $ and 
atmospheric turbulence in the cloud formation and the distribution of vortices and eddies in the form
of the logarithmic spiral with the golden mean winding number havs been analyzed by Selvam [2] 

\smallskip
Montgomery has shown [3] that the two-level correlation function of the
imaginary parts of the zeroes of zeta is exactly the same expression
found by Wigner and Dyson using Random Matrices techniques : the two-level
spectral density correlation function of the Brownian-like
discrete level statistical dynamics associated with the random matrix
model of a Gaussian Unitary Ensemble ( GUE) turned out to be :
$$ 1 - [{sin (\pi x) \over \pi x }] ^2. \eqno (1) $$
The function $sin (\pi x) \over \pi x$ appears very naturally  as well in the
self similarity of the iterated symbolic dynamics of the binary Fibonacci
(rabbit) sequence of numbers .  It is the Fourier amplitude 
spectrum of the iterated binary sequence and the golden mean plays a fundamental role in generating a
long range but aperiodic order , a one dimesnional quasi-crystal,  see the book by
Schroeder [20] . 
The golden mean is the $n = \infty$ limit of
the ratio of two successive Fibonacci numbers. 

Physically the relation (1)
means that there is a " repulsion"  of the energy levels; i.e. the probability of two energy levels 
being very close becomes very small. 
For signals of Chaos in M(atrix)  theory ( Yang- Mills ) and its deep relation to
the holographic properties in string and $M$ theory we refer to the
important work of Volovich et al [4]. Since string theory has a deep
connection to the statistical properties of random surfaces, index theory
for fractal $p$-branes in Cantorian fractal spacetime was considered by
the author and Mahecha [16] in connection to the Rieman zeta function.
The spectrum of drums ( membranes) with fractal boundaries bears deep
relations to the zeta function for those fractal strings that are not Minkowski measurable 
[18].  A random walk based on $p$-Adic numbers has been studied by  Albeverio and Karkowski [25] 

In this letter we will combine all these
ideas with the fundamental inclusion of supersymmetry. Parisi and Sourlas
[5] discovered in the late 70's that there is a hidden supersymmetry in
classical stochastic differential equations. The existence of stationary
solutions to the Fokker-Planck equation associated with the stochastic
Langevin equation can be reformulated in terms of an unbroken supersymmetric
(SUSY) Quantum Mechanical Model . More precisely, with an imaginary time
Schroedinger equation : a diffusion equation involving a dual diffusion
process forward and backward in time. Nagasawa , Chapline [6,7] have used
these ideas to reformulate QM as information fusion. Ord and Nottale [8, 9]
following the path pioneered by Feynman path's integral formulation of QM,
have shown that the QM equations can be understood from an underlying
fractal dynamics of a particle zig-zagging back and forth in spacetime
spanning a fractal trajectory ; i.e. a particle undergoing a random walk or
Brownian Motion.  Armitage [20] proposed long ago a random walk ( diffusion ) 
approximation to 
the Riemann Hypothesis : a random walk approach to the Ornstein-Uhlenbeck process 
( or Fokker-Planck equation) 
to exhibit a polynomial whose zeroes, under a suitable limiting process, 
ought to be zeroes of the Riemann zeta. 
El Naschie [ 10 ] has suggested also that this dual diffusion
process could be the clue to prove the Riemann conjecture. 

We will propose here the physical dynamical model that
furnishes , in principle, once a solution of the quantum inverse
scattering problem exists, the sought-after Hilbert-Polya operator which
yields the zeroes of zeta . If a solution to this quantum inverse scattering
problem exists this could be instrumental in proving the Riemann Conjecture.
We will make use of all of these ideas of random matrix theory , Brownian motion,
random walk, fractals, quantum chaos, stochastic dynamics ...within the
framework of the supersymmetric QM model associated with the Langevin
dynamics and the Fokker-Planck equation of a particle moving in a randomly
fluctuating medium; i.e. $noise$ due to the random fluctuations of the
infinite array of particles (atoms ) located along the one dimensional
quasi periodic crystal.

By random fluctuations we mean those fluctuations with respect to their
equilibrium configurations which, for example, could be assumed to be the
locations of the integers numbers. Because the prime numbers distribution is connected 
to the distribution of the zeroes of zeta,
the main idea of this work is to recast this physical problem in terms of the scattering of the particle by 
scattering centers situated at the prime numbers,  and in this fashion we
have effectively a random process with an underlying Parisi-Sourlas hidden
supersymmetry , and hence , a well defined SUSY QM problem.

Watkins [11 ] has also suggested that an infinite array of (charged )
particles located at the positions of the prime numbers could be relevant
in describing the physical system which provides the evolution dynamics
linked to the zeroes of zeta. Pitkanen [12] has refined Riemann's
conjecture within the language of p-Adic numbers by constraining the
imaginary parts of the zeros of zeta to be members of complex rational
Pythagorean phases and Berry and Keating [ 13 ] have proposed that the
SUSY QM Hamiltonian :
$$ H = xp - i = Q^2 \eqno (2) $$
is relevant to generate the imaginary parts of the zeroes.

The imaginary time Schroedinger equation (diffusion equation ) that we
propose is :
$$ - D {\partial \over\partial t} K_{\pm} ( x, t) = H_{\pm} K_(x, t).
\eqno (3) $$
where $ D$ is the diffusion constant which can be set to unity , in the
same way that one can set $ \hbar = m = 1$ where $m$ is the particle's
mass subject to the random walk .

The $isospectral$ partner Hamiltonians,
$ H_+, H_-$ are respectively :
$$H_{\pm} = - {D^2\over 2} {\partial^2 \over \partial x^2} + { 1\over 2}
\Phi^2 \pm { D\over 2} {\partial \over \partial x} \Phi. \eqno (4) $$
the transition- probability density solution of the Fokker-Plank equation ,
$ m{\pm} (x, x_o, t)$ , for the particle arriving at $x$ , in a given time
$t$ , after having started at $x_o$ is :
$$ m^{\pm} (x, x_o, t) = exp ~[ - {1\over D} ( U_{\pm} (x) - U_{\pm} (x_o))
 ] ~K_{\pm} (x, t). \eqno (5)$$

The Fokker-Planck equation obeyed by the transition- probability density
is :
$${\partial \over \partial t} m(x, x_o, t) = { D \over 2} {\partial^2
\over \partial x^2} m (x, x_o, t) + {\partial \over \partial x }\Phi (x)
m(x, x_o, t). \eqno (6a)$$
and the associated Langevin dynamical equation :
$${\partial x \over dt} = F (x) + \xi (t). \eqno (6b) $$
$ F = \Phi (x) $ is the drift momentum experienced by the particle . The
quantity $\xi (t) $ is the $noise$ term due to the random fluctuations of
the medium in which the particle is immersed. The drift potential $U(x)$
associated with the stochastic Langevin equation is defined to be :
$$ U_{\pm} (x ) = - (\pm) \int_0^x dz \Phi (z ) \eqno (7) $$
$\Phi (x)$ is precisely the SUSY QM potential as we shall see below. The
two partners $isospectral$ ( same eigenvalues ) Hamiltonians can be
$factorized$ :
$$H_{+} = {1\over 2} ( D {\partial \over \partial x} + \Phi (x))
( D {\partial \over \partial x} - \Phi (x)) = {\cal L}^{-}{\cal L}^+ .
\eqno (8) $$
$$H_{-} = {1\over 2} ( D {\partial \over \partial x} - \Phi (x))
( D {\partial \over \partial x} + \Phi (x)) = {\cal L}^{+}{\cal L}^{-}
\eqno (9)$$
If SUSY is unbroken there is a $zero$ eigenvalue $\lambda_o = 0$ whose
eigenfunction corresponding to the $H_-$ Hamiltonian is the ground state :
$$ \Psi^{-}_0 (x) = C e^ {- {1\over D} \int_0^x dz \Phi (z)}.\eqno(10)$$
$C$ is a normalization constant.

Notice that the random " momentum" term $\xi(t)$ appearing in
Langevin's equation can be simply recast in terms of the other quantities
as :
$$ {\partial x \over dt} = F (x) + \xi (t) \Rightarrow{\partial x\over dt}
- F(x)= \xi (t) .\eqno(11) $$
which in essence means that the random potential term (in units of
$\hbar =m =1$) is :
 $$ \xi (t) = p - F(x) = p + \Phi (x) . \eqno (12) $$
which is just
the $\cal L^{-}$ operator used to factorize the Hamiltonian  $H_{-}$. One has
two random potential terms : $\xi^{\pm} (t)$ corresponding to the two
operators $\cal L^{-}, \cal L^{+}$ associated with the two isospectral
Hamiltonian partners $ H_{\pm}$.

An immediate question soon arises. Since the imaginary parts of the zeroes
of zeta do $not$ start at zero , the first zero begins at
$y = 14.1347.....$ , how can we reconcile the fact that the ground state
eigenfunction has zero for eigenvalue ( by virtue of SUSY) ? . The answer
to this question has been discussed  by Pitkanen [ 12]  using theoretical arguments involving 
$p$-adic numbers. This was based on earlier work by
Julia [22]   who constructed the fermionic version of
the zeta function and had shown that the bosonic and fermionic zeta function can be recast as
partition functions of systems of p-adic bosonic/fermionic oscillators in
a thermal bath of temperature $T$. The frequencies of those oscillators is
$ log ~p$, for $ p =2,3,5...$ a prime number and the inverse temperature
$1/T$ corresponds to the $z$ coordinate present in the $\zeta (z)$ where
$real~ z >1$. The pole at $ z =1$ naturally corresponds to the limiting
Hagedorn temperature.

By virtue of SUSY the zeroes of the fermionic zeta function coincided
precisely with the zeroes of the bosonic zeta with the fundamental
difference that the fermionic zeta had an additional zero precisely at
$ z = 1/2 = 1/2 + Oi $ ; i.e the imaginary part of the first zero of the
fermionic partition function is precisely $zero$ !.

Therefore, in this SUSY QM model we naturally should expect to have a zero
eigenvalue associated with the supersymmetric ground state. Such a ground state
does $not$ break SUSY and ensures that the associated Fokker-Planck equation
has a $stationary$ solution in the $ t = \infty$ limit ( equilibrium
configuration is attained at $ t = \infty$ ). Such stationary solution
is given precisely by the modulo-squared of the ground state solution to the
SUSY QM model [5]:
$$ \lim ~  m^{-}_{t \rightarrow \infty} ( x, x_o = 0 , t) = P(x) = |\Psi_0^-(x)|
^2 =C^2 exp~[ - { 2 \over D} U_-(x)]. \eqno (13)$$
Notice that the ground state solution is explicitly given in terms of the
potential function$ U_- (x) $ given by the integral of the SUSY potential
$ \Phi (x)$ ; i.e .${\partial \over \partial x} (- U_{\pm} (x)) = \pm
\Phi (x)$.

One should notice that the $ordinary$ harmonic oscillator corresponds
roughly speaking to the case : $ \Phi \sim x $ so the operators
${\cal L}^+, {\cal L}^-$ match the raising and lowering operators in this
restricted case. However this is a very special case and $not$ the SUSY
QM model studied here. The physical model we are studying is $not$ an
ordinary $real$ harmonic oscillator but instead it is a $p$-Adic one
related to Pitkanen's original $p$-adic bosonic/fermionic harmonic
oscillator formulation of the zeta function as $p$-adic partition
functions.

In principle, if one has the list of the imaginary parts of $all$ the
zeroes of zeta one can $equate$ them to the infinite number of eigenvalues:
 $$ \lambda_o = 0, \lambda_n = \lambda^-_n = \lambda^+_n ~~
 n =1,2,3..... \eqno (14) $$
The main problem then is to find the SUSY potential $\Phi (x)$
associated with the zeroes of the bosonic/fermionic zeta. This
would require solving the inverse quantum scattering method (that
gave rise to quantum groups). To do this is a formidable task
since it requires to have the list of the $infinite$ number of
zeroes to begin with and then to solve the inverse scattering
method problem. Wavelet analysis is very suitable for solving 
inverse scattering methods. Not surprisingly, Kozyrev has given
convincing arguments that wavelet analysis is nothing but
$p$-adic harmonic analysis [17].

What type of SUSY potential do
we expect to get ? Is it related to the scattering of the
particle by an infinite array of atoms located at the prime
numbers ? Is it related to the Coulomb potential felt by the
particle due to the infinite array of charges located at the
prime numbers ? Is it related to a chaotic one-dimensional
billiard ball where the bouncing (scattering ) back and forth
from an infinite array of obstacles located at the prime numbers
? Is it just the $p$-adic stochastic Brownian motion modeled by
Pitkanen's $p$-adic bosonic/fermionic oscillators ? Since the
properties of the zeta function are associated with the
distribution of primes numbers it is sensible to pose this list of 
questions. We will try to answer these questions shortly.

The
ordinary QM potential associated with the SUSY $\Phi (x)$
potential is defined as $$ V_{\pm} (x) = {1\over 2}\Phi^2 (x) \pm
{D\over 2} {\partial \over \partial x} \Phi (x). \eqno (15)$$ The
potential $ V( x)$ should be symmetric (to preserve supersymmetry)
under the exchange $ x \rightarrow - x$ and this entails that the
SUSY potential $\Phi (x) $ has to be an odd function : $ \Phi (-
x) = - \Phi (x) $.

Clearly since we do not have at hand the
algorithm to generate all the zeros of zeta ( nor the prime
numbers ) one cannot write explicitly the ansatz for the
potential nor solve the inverse scattering problem (which would
yield the form of the potential). Nevertheless , Euler was
confronted with a similar problem of finding out all the prime
numbers when he wrote the adelic product formula of the Riemann
zeta which relates an infinite summation over the integers to an
infinite product over the primes : $$ For ~ Real ~ z > 1 ~~\zeta
(z) \equiv \sum^\infty {1 \over n^z} = \prod _{p} (1 - p^ {- z})
^{-1}.\eqno(16)$$ Euler's formula can be derived simply by
writing any integer as a product of powers of primes and using
the summation formula for a geometric series with growth
parameter  $p^{-z} $. Euler's formula is a simple proof of why
the number of primes  is  infinite. The product is an infinite
product over $all$ primes. Despite the fact that we do not have
the list of $all$ the prime numbers nor an algorithm to generate
them this does nor prevent
us from evaluating such an infinite product; i.e in computing the
value of the zeta function by performing the sum of the Dirichlet
series over all integers!

This will provide us with the
fundamental clue for writing an ansatz for the SUSY potential
$\Phi ( x) $ and its associated  potentials $V_{\pm}(x)$ giving
us the SUSY QM model which yields all the imaginary parts of the zeroes of the
zeta function as the eigenvalues of such SUSY QM model. The
reader could ask why should we go to all this trouble and take a
tortuous route of writing down the SUSY QM model instead of
solving directly the ordinary QM inverse scattering problem ; i.e
. finding the ordinary potential $ V_{\pm} (x) $ of an ordinary QM
problem ?

If one followed such procedure one would lose the deep
underlying stochastic dynamics of the problem. One would have not
discovered the underlying hidden supersymmetry associated with
the stochastic Langevin dynamics; nor its associated
Fokker-Planck equation; nor be able to notice that the ground
state is supersymmetric and its eigenvalue is precise $zero$ ;
nor to construct the ground state solution explicitly in terms of
the SUSY potential $\Phi (x)$ as shown in Eq.(13 ). Also one
would  fail to notice the crucial $factorization $ properties
of the Hamiltonian, and that the potential $ V_{\pm} (x)$ must be
symmetric with respect to the origin while the SUSY potential
$\Phi (x)$ is antisymmetric,.....etc. The SUSY QM is very restricted
that narrows down the inverse scattering problem. The simplest
analogy one can give is that of a person who fails to recognize
the sine function because instead there is the infinite Taylor
expansion of the sine function.

The crux of this work is to write
down the form of the potential associated with the SUSY QM
model. The only assumption of this work is based in writing the sought-after SUSY potential in
terms of an infinite product, similar to the Euler Adelic product
form of the zeta and to well known relation between the $sums$ of
gamma functions in terms of $products$ of zeta functions present
in the scattering formulae of $p$-adic open strings [4, 14 ]. We
proceed as follows. $$ \Phi (x) \equiv\sum_n V (| x - x_n|) =
\prod_{p} W (x_p) \eqno (17) $$ where $x_p
\equiv p^ {- x} $ and $ W = W( x_p)$ is a function to be
determined. This is our ansatz for the form of the SUSY potential. We shall
call this ansatz for the potential the adelic potential condition since the
product is taken over all the primes. All we are assuming is that a potential of this form can be found. 

The zeta function has a
similar form ( although it is not symmetric with respect to the
origin ): one has an infinite summation over all the integers of
the series $ n^{- z } $ ( playing the role of the potential)
expressed as an infinite product over all the primes of functions
of $p^ {-z} $. We are just recasting the potential felt by the
particle due to the infinite interactions with the objects
situated at $ x_n $ , an infinite sum of terms, in    terms of the
infinite product over all the primes of functions of $ x_p \equiv
p^{-x }$. We emphasize that what is equal to the imaginary parts
of the zeroes of zeta are the $eigenvalues $ $\lambda_n$ of the
SUSY QM model. We have just recast the infinite numerical input
parameters $ x_n$ of the potential in terms of the location of
the infinite number of primes .

For example, based on the Euler
adelic formula for the zeta function, one could have chosen the
potential to have  precisely the zeta function
form : $$For ~ x >0. ~~\Phi (x) = \prod_{p} W (x_p) =
\prod_{p}(1- x_p )^{-1} .\eqno(18)$$
\centerline{$x_p
\equiv p^{-x}, \Phi(x)=-\Phi(-x)$}
\smallskip

 where $ x$ is the
location of the particle executing the $p$-adic stochastic
Brownian motion . For applications of $p$-adic numbers in physics
we refer to [4,14]. Due to the antisymmetry requirement of
$\Phi(x) $ , the SUSY potential for $x< 0$ must be taken to be an
exact mirror copy of the $ x > 0$ region to ensure that
supersymmetry is unbroken. As a result, the partner potentials
$V_{\pm}$ are \footnote {To ensure that $\Phi(-x)=-\Phi(x)$ one
must add a constant such that $\Phi(0)=0$}: $$ V_{\pm} (x) =
{1\over 2} \Phi^2 \pm { D \over 2} {\partial \over \partial
x}\Phi (x). \eqno (19) $$

But how can we be so sure that the eigenvalues will be precisely
equal to the imaginary parts of the zeroes of zeta ? This would have been an
$amazing$ coincidence ! For this reason we must have an unknown
function $ W = W(x_p)$ to be $determined$ by solving the quantum
inverse scattering method. Assuming that the eigenvalues are
precisely the imaginary parts of the zeroes of zeta , in
principle , we have defined the quantum inverse scattering
problem . Using wavelet analysis or $p$-adic harmonic analysis
one could find the SUSY potential $\Phi(x) = \prod_{p} W (x_p)$
where $ x_p\equiv p^{-x } $ , where $ x$ is the location of the
particle. How does one achieve such a numerical feat ? One could
expand the function $ W(x_p)$ in Taylor series assuming that the
potential is analytic, except at some points representing the
location of the infinite array of particles, obstacles of the
chaotic one-dimensional billiard, or the atoms ( scattering
centers ) of the one dimensional quasi periodic crystal . The
adelic condition for the SUSY potential becomes then :
$$\begin{array}{lc} For~~x>0
~~\Phi(x)= \prod_{p} W (x_p) = \prod_{p} ~\sum_{n} a_n ( x_p)^n\\
\\
=\prod_{p} ~ \sum_{n} a_n [ p^ {-x}]^n =
 \prod_{p} \sum_{n} a_n p^ {- n x}.
\end{array}
 \eqno (20a) $$
where $\Phi(x)=-\Phi(-x)$  must also be imposed. To ensure that $\Phi ( 0 ) = 0$ one must add a suitable constant if necessary .

Recasting the Taylor series as a Dirichlet series by simply rewriting :
$$ \begin{array}{lc}
p^ {-nx } = e^{ -x nlog
p } = e^{ i^2 x n log~p } \rightarrow\\ \Phi(x)
= \prod_{p} \sum_n a_n cos
~ [i x (  log~p^n)] +i a_n sin~ [ i x ( log~p^n) ].
\end{array}
\eqno (20b) $$
(where once again $\Phi(x)=-\Phi(-x))$ allows us to expand the adelic potential in terms of an 
$infinite$
collection of $p$-adic harmonic oscillators with fundamental
imaginary frequencies $ \omega_p = i log~ p $ and whose harmonics
are suitable $powers$ of the fundamental frequencies : $
\omega_{p, n } = i log ~ p^n = i n~ log~ p $. One can recast the imaginary argument trigonometric
functions in terms of hyperbolic functions if one wishes. Hyperbolic functions ( potentials ) 
are very natural in SUSY QM models. A whole class of solvable potentials, like the shape-invariant partner 
potentials,  have been discussed amply in the book by G. Junker [ 5 ].  

We have then recast the quantum inverse scattering problem ( solving for the SUSY potential)  as the problem
of solving for the amplitudes $ a_n$ (and phases ) of the (imaginary frequencies
) $p$-adic harmonic oscillators by simply writing the adelic condition on the SUSY 
potential in terms of a $p$-adic Fourier expansion ($p$-adic
harmonic analysis ) . This is attained by means of performing the
usual Wick rotation in Euclidean QFT : $ \omega \rightarrow i
\omega $. One could Wick rotate the imaginary time Schroedinger
equation (a diffusion equation ) to an ordinary Schroedinger
equation by the usual Wick rotation trick $ t \rightarrow i t $ .
Our $p$-Adic Fourier expansion condition on the SUSY potential coincides with Pitkanen's ideas about 
the zeta function being the partition function of the adelic
ensemble of an infinite system of $p$-adic oscillators with
fundamental frequencies $ \omega_p = log ~ p$, with the only
difference that in our case we are performing the Wick rotation
of those frequencies.

The infinite unknown amplitude coefficients
$ a_n$ will be determined numerically by solving the inverse
quantum scattering problem in terms of the eigenvalues of the
SUSY QM model = imaginary parts of the zeroes of zeta. A. Odlyzko
[15] has computed the first $ 10^{20}$ (or more) zeroes of zeta.
Having a list of $ 10^{20}$ zeroes should be enough data points
to find the first $10^{20}$ numerical coefficients $a_n$
appearing in the Taylor expansion of the potential function $\Phi
(x) = \prod_{p} W(x_p)$ that is being determined via inverse
quantum scattering methods associated with this SUSY QM model
that links stochastic dynamics , supersymmetry, chaos, ....etc. to the
zeroes of the Riemann zeta function.

To summarize, if one can find
a solution of the inverse scattering problem that determines the
(symmetric) potential $ V_{\pm} (x )$ then one can

 1) propose the candidate for the 
 Hilbert-Polya operator in the following form:

 $${\cal H} =
1/2 ( H_- H_ + + H_+ H_-) + 1/4 .
\eqno (21 )$$
 and

            2) postulate that the eigenfunctions $\Psi_n$ of the
Hamiltonian containing quartic derivative $\cal{H}$ is the $fusion$
( or convolution) of two eigenfunctions $\Psi^+_{n-1}$ and $\Psi^-_{n}$.
The $ordinary$ product will not be suitable, as can be verified by simple
inspection. The fusion rules of this type have been widely used in conformal field
theories and in string theory.

The fused quartic derivative Hamiltonian operator is automatically self-adjoint as a result of
the $fusion$ of the  $self-adjoint$  isospectral Hamiltonians $ H_+, H_-$ which
characterize the two " dual " Nagasawa's diffusion equations [10]
and its $real$ eigenvalues are the product of the nontrivial
zeroes of the Riemann zeta function and their complex
conjugates :
  $$ {\cal H} \Psi_n = (1/4 + \lambda^2_n) \Psi_n = (1/2 + i \lambda_n )
  (1/2 -i\lambda_n ) \Psi_n.\eqno (22)$$.
Notice that the value $ n = 0$ is not included since $ \Psi^+ _{-1} $ is
$not$ defined and the operator ${\cal H} $ is $quartic$ in derivatives.
Could this candidate for the Hilbert-Polya operator be instrumental in proving
the Riemann conjecture ? The Eguchi-Schild action for the string is the square of the  
Poisson bracket of the string embedding coordinates with respect to the worldsheet
variables.  The momentum conjugate to the Eguchi-Schild holographic area variables is called the area-momentum. 
The square of the area-momentum is in this sense quartic in derivatives.  

The fusion or convolution product of the two eigenfunctions of $ H_{\pm} $
can be found by referring to the Fourier transform: the Fourier
transform of an ordinary product equals the convolution product of their
Fourier transforms . Hence the eigenfunctions of ${\cal H} $ can be written,
by denoting $ F, F^{ -1} $ the Fourier transform and its inverse :
$$ \Psi_n (x) = F^{-1} [ F(\Psi^+_{n-1} ) * F( \Psi^-_{n}) ] .~~ n =1,2,3,...
\eqno (23)$$
As expected, the eigenfunction of the fused Hamiltonian is $not$ the naive
product of the eigenfunctions of their constituents.

The $1/4$ coefficient present in the eigenvalues of the fused-Hamiltonian
operator is intrinsically related to the real part of the zeroes of zeta
$ (1/2 + i\lambda_n) (1/2 + i \lambda_n). $ The interpretation of the
$1/4$ coefficient appearing in the fused-Hamiltonan is as an $additive$
constant due to normal ordering ambiguities in QFT , like a zero point energy of the ordinary Harmonic oscillator.
>From the conformal field theory and string theory point of view one
constructs unitary irreducible highest-weight representations of the
Virasoro algebra for suitable values of the central charges and weights
associated with the ground states , $ c$ and $h$ respectively. It is very
plausible that supersymmetry, superconformal invariance and representation theory may select and fix
uniquely the value of $1/4 $ which then would be an elegant proof of the
Riemman conjecture. 

Pitkanen [12] has used conformal field theory
arguments to refine the Riemman conjecture : the imaginary parts of
the zeroes correspond to complex rational Pythagorean phases : $ p^ {iy}$ . Another way of rephrasing this is by saying that 
the Pythagorean phases correspond to 
the rational points solutions of the phase space energy surface of a harmonic oscillator 
$ P^2 + X^2 = 1$.  Just  
recently he has proposed a non-Hermitean Hilbert-Polya operator of the type $ L_o + V $ 
whose eigenvalues are $1/2 + i y $ using superconformal field arguments. The inner product 
of two eigenfunctions was re-expressed in terms of the zeta function and the orthogonality 
of states amounted to the vanishing of zeta.  $L_0$ is the zero mode of the Virasoro scaling generator 
in the complex plane and $V$ is a suitable potential.
A coherent states interpretation may allow to express the inner product of states in terms of a non-trivial 
metric with varying scalar curvature.  A  constant negative scalar curvature belongs to a hyperbolic geometry, like the Poincare
disc in the complex plane.  
The Selberg zeta function allows to count the primitive periodic orbits ( geodesics )  of a chaotic billiard 
on a hyperbolic plane.  The proyect will be in finding out whether or not the Riemann zeta function can be extracted in 
this physical model using conherent states methods.  The location of the zeroes of zeta will amount to a 
destructive interference between the holomorphic modes ( right-moving ) and the anti-holomorphic modes ( left moving ) associated
with the coherent states. 

Fractal $p$-branes in Cantorian-fractal spacetime and its relation to the
zeta function were considered by the author and Mahecha [ 16]. The role of fractal strings 
and the zeroes of zeta has appeared
in the book by Lapidus and van Frankenhuysen [ 18] . Combining $p$-adic
numbers and fractals we arrive at the notion of $ p$-adic fractal strings.
The fundamental question to ask would be how to establish a one-to-one
correspondence between the zeroes of zeta and the spectrum of $ p$-adic
fractal strings. This would mean an establishment of a relation between
the exact location of the $poles$ of the scattering amplitudes of $p$-adic
fractal strings (a generalization of the Veneziano formula in terms of
Euler gamma functions) to the exact location of the zeroes of zeta,  i.e.
a one -to-one correspondence between the Regge trajectories  in the
complex angular momentum plane  and the spectrum of the $p$-adic fractal
strings with the zeroes of zeta .

Since $p$-adic topology is the topology of Cantorian-fractal spacetime
it is not surprising that the Golden Mean will play a fundamental role
[2, 10, 16]. $p$-adic fractals have been discussed in full detail by
Pitkannen [12]. $p$-Adic Fractals are roughly speaking just $fuzzy$
fractals [19], as they should be, since Cantorian-fractal spacetime
involves a von Neumann's noncommutative $pointless$ geometry .
Wavelet analysis = $p$-adic Harmonic analysis must play a fundamental role
[17] in the classification of such spectrum. After all, the scattering of
a particle off a $p$-adic fractal string should be another way to look at
the $p$-adic stochastic motion discussed in this work.

\bigskip

\centerline{\bf Conclusion}
\bigskip
 
We have been able to construct (assuming the quantum inverse
scattering problem has a solution ) the operator that yields the zeroes of
the Riemann zeta function by defining explicitly the SUSY QM model
associated with the $p$-adic stochastic dynamics of a particle undergoing
a Brownian random walk . The zig-zagging occurs after collisions with an
infinite array of scattering centers that fluctuate $randomly$ .  We argued why this physical system
can be $reformulated$ as the scatterings of the particle about the
infinite locations of the prime numbers positions.  Assuming that the SUSY potential 
admits a $p$-adic Fourier decomposition, we have
reformulated such $p$-adic stochastic process which has an underlying
hidden Parisi-Sourlas supersymmetry as the effective Brownian motion of a test 
particle moving in a background of a thermal gas of photons ; i.e  the quanta excitations of an  
infinite collection of $p$-adic harmonic oscillators
with fundamental imaginary frequencies given by $\omega_p = i log~ p$ and whose
harmonics are $ \omega_{p, n} = i log~ p^n$. This is what we called the adelic ansatz condition for the
SUSY potential and that allowed us to determine a one-to-one correspondence between
the amplitudes of oscillations $a_n$ ( and phases ) with the imaginary parts of the 
zeroes of zeta $ \lambda_n$ , after solving the inverse scattering
problem. $p$-adic fractal strings and their spectrum may establish a
one-to-one correspondence between the poles of their scattering amplitudes
and the zeroes of zeta.
\bigskip

\centerline {\bf Acknowledgements }

\smallskip
We are very grateful to A. Granik for his help in preparing this
manuscript and for many discussions. We also wish to thank M. Pitkannen,
M. S. El Naschie, J. Mahecha for their correspondence.  Special thanks to M. Watkins for critical reading of 
the manuscript and for  corrections and references  and to J. A. Boedo for assistance.

\bigskip

\centerline {\bf References}

\smallskip
1- M. L Metha : "Random Matrices" Academic Press , New York 1991.
\smallskip

A. Bandey :" Brownian motion of Discrete Spectra " J. Chaos, Solitons and
Fractals vol 5, no.7( 1995) 1275.
\smallskip

2- A. M. Selvam :" Wave-Particle Duality in Cloud Formation" . 
"Universal Quantification for Self Organized Criticality in
Atmospherical Flows".  http://www.geocities.com/amselvam /pns97.html

\smallskip
3- H. Montgomery : "Distribution of the Zeroes of the Riemann Zeta
Function"  Proceedings Int. Cong. Math, Vancouver 1974, Vol 1, 379-381.

\smallskip
4. I. Aref'eva, P. Medvedev, 0. Rythchkov, I. Volovich :
"Chaos in Matrix) Theory" J. Chaos, Solitons and Fractals vol. 10 ,
nos 2-3 (1999 );

V. Vladimorov, I. Volovich, E. Zelenov : "p-Adic Analysis in Mathematical
Physics" World Scientific , Singapore 1994.

\smallskip

5- J.Parisi and N.Sourlas, "Supersymmetric field theories and stochastic
   differential equations", Nucl.Phys B {\bf 206} (1982) 321;

G.Junker, "Supersymmetric methods in quantum and statistical mechanics",
Springer-Verlag 1996

K. Efetov,"Supersymmetry in Disorder and Chaos" Cambridge University
Press , 1999;

K. Nakamura,"Quantum Chaos" Cambridge University Press 1995;

\smallskip

6-G. Chapline,"Quantum Mechanics as Information Fusion ",quant-ph/9912019

\smallskip

7-M. Nagasawa, Special issue of the J. of Chaos, Solitons and Fractals,
vol. 7 ,no. 5 (1996 ) on "Chaos, Information and Diffusion in Quantum
Physics".

\smallskip

8- L. Nottale , "The Scale Relativity Program " J. of Chaos, Solitons and
Fractals10, nos 2-3 (1999) 459;

L. Nottale, "La Relativite dans Tous ses Etats", Hachette Literature
Paris,1998.

\smallskip

9- G. Ord, " The Spiral Gravity Model" J. of Chaos, Solitons and
Fractals vol. 10 , nos 2-3 (1999 ) 499.

\smallskip

10- M. S. El Naschie, "Remarks on Superstrings, Fractal Gravity,
Nagasawa's Diffusion and Cantorian Fractal Spacetime", J. of Chaos,
Solitons and Fractals vol.8, no. 11 (1997)1873

\smallskip

11- M. Watkins,"Prime Evolution Notes" http://www.maths.ex.ac.uk/$\sim$mwatkins
/zeta/evolutionnotes.htm

\smallskip

12- M/ Pitkanen, "Topological Geometrodynamics " Book on line :
http: //www.physics.helsinki.fi/~matpitka/ tgd.htmf

\smallskip

13-M. Berry and J. Keating,
"The Riemann zeroes and eigenvalue asymptotics" SIAM Review {\bf 41}
no. 2 (1999 ) 236-266.

\smallskip

14- L. Brekke and P. G.O. Freund,"p-Adic Numbers in Physics"
Phys. Reports vol. 233 no. 1 (1993).

\smallskip

15-. A. Odlyzko,"Supercomputers and the Riemann zeta function" Proc. of the fourth Int. conference on 
Supercomputing. L P Kartashev
and S. I. Kartashev (eds). International Supercomputer Institute 1989,
348-352.

\smallskip

16- C. Castro, J. Mahecha, "Comments on the Riemann Conjecture and Index
Theory on Cantorian Fractal Spacetime" hep-th/0009014 v2,
to be published in the J. Chaos, Solitons and Fractals.

\smallskip

17- S. Kozyrev,"Wavelets Analysis as p-Adic Harmonic Analysis"
math.ph / 0012019

\smallskip

18- M. Lapidus and M. van Frankenhuysen, "Fractal geometry and Number theory:
fractal strings and zeroes of zeta functions". Birkhauser ( 2000 ). 

\smallskip

19- A.Granik, Private communication

20-Schroeder : " Chaos, Fractals and Power laws "

21- J. V. Armitage : " The Riemann hypothesis and the Hamiltonian of a quantum mechanical system "
 In Number Theory and Dynamical Systems " edited by M. M. Dodson and J. Vickers, London Mathematical 
Lecture Notes series 134,  Cambridge University Press ( 1989) pages 153-172.

22- B. Julia : See http://www.maths.ex.ac.uk/$\sim$mwatkins/zeta/physics2.htm

23. M. PItkanen : " Proof of Riemann Hypothesis  " math.GM/0102031 

24.  W.  Karkowski, R. Vilela Mendes : Journal. Math. Physics {\bf 35} ( 1994) 4637. 

25 - S. Albeverio, W. Karkowski : " A Random Walk on p-Adics " Stochastic Processes and Applications 
{\bf 53} ( 1994) 1-22.

\end{document}